# Product number counting statistics from stochastic bursting birth-death processes

Seong Jun Park[1], Jaeyoung Sung*[1,2]

[1] National CRI-Center for Chemical Dynamics in Living Cells, Chung-Ang University, Seoul 06974, Korea.

[2] Department of Chemistry, Chung-Ang University, Seoul 06974, Korea.

**Abstract**

Bursting and non-renewal processes are common phenomena in birth-death process, yet no theory can quantitatively describe a non-renewal birth process with bursting. Here, we present a theoretical model that yields the product number counting statistics of product creation occurring in bursts and of a non-renewal creation process. When product creation is a stationary process, our model confirms that product number fluctuation decreases with an increase in the product lifetime fluctuation, originating from the non-Poisson degradation dynamics, a result obtained in previous work. Our model additionally demonstrates that the dependence of product number fluctuation on product lifetime fluctuation varies with time, when product creation is a non-stationary process. We find that bursting increases product number fluctuation, compared to birth-processes without bursting. At time zero, in a burst-less birth process, product number fluctuation is unsurprisingly found to be zero, but we discover that, in a bulk creation process characterized by bursting, product number fluctuation is a finite value at time zero. The analytic expressions we obtain are applicable to many fields related to the study system population, such as queueing models and gene expression.

The birth-death process is one of the most commonly occurring phenomena in nature and the main interest of queueing theory, population dynamics, and other related fields [1-4]. As such, birth-death models have been studied extensively for decades [5-7]. One such example can be found in Lim. *et. al* [8], where the researchers introduce a new type of stochastic kinetics for a chemical reaction whose rate coefficient is a stochastic variable; this process is called a vibrant reaction process and capable of representing both renewal and non-renewal processes. Furthermore, the Chemical Fluctuation Theorem (CFT) governing the vibrant-birth process and renewal-death process was shown to successfully and quantitatively provide an accurate description of gene expression [9].

Bursting, where many product molecules are created at once, also commonly occurs during the birth process. This is evidenced by single molecule measurements, where examples of bursting can be seen in the production of biomolecules such as mRNA and protein [10-13]. To analyze experimental studies with bursting birth-processes, Tao. *et. al* [14] used a $GI^x/G/\infty$ system from queueing theory [7]. According to queueing theory literature, a birth-death process is described by a series of symbols and slashes such as A/B/C, in which A indicates the birth-time distribution, B the death-time distribution, C the number of death channels. In a $GI^x/G/\infty$ system, G refers to the general waiting-time distribution and $I^x$ indicates that products are produced in batches of random size X, where X is a discrete random variable greater than or equal to unity.

The product counting statistics from the $GI^x/G/\infty$ system can be derived by considering both birth and death processes as renewal processes. However, experiments have shown that, in enzymatic reactions and gene expression, product birth rates are random variables, indicating that product birth may not be a renewal process [15-17].

In this work, we extend the CFT to incorporate a bursting birth process, where the number of products can be created of any size. We obtain useful, analytic expressions of the product number of a bursting product creation process for the first and second moments. With these results, we then investigate the counting statistics of product molecules that undergo vibrant birth processes, occurring in batches of a random size, and renewal death processes.

Let us consider the product in Fig. 1(a), the underlying reaction scheme considered in the current work. The wavy arrow in Fig 1(a) represents a vibrant-birth process. The analytic expressions for the mean and the variance of the product number are respectively given by Eqs. (1) and (2) below:

$$\langle n(t) \rangle = \langle b \rangle \int_0^t d\tau \langle R(\tau) \rangle S(t-\tau) \tag{1}$$

$$\sigma_n^2(t) = \langle n(t) \rangle + \left( \langle b^2 \rangle - \langle b \rangle \right) \int_0^t d\tau \langle R(\tau) \rangle S(t-\tau)^2 \\ + 2\langle b \rangle^2 \int_0^t d\tau_2 \int_0^{\tau_2} d\tau_1 \langle \delta R(\tau_2) \delta R(\tau_1) \rangle S(t-\tau_1) S(t-\tau_2) \tag{2}$$

In the above equations, $\langle R(t) \rangle$ is the mean product-birth rate, $S(t)$ denotes the survival probability of product molecules, $\langle b^m \rangle$ is the m-th moment of burst size $b$, and $\langle \delta R(t_2) \delta R(t_1) \rangle$ is the time correlation function (TCF) of the birth rate in the presence of birth rate fluctuation, $\delta R(t) = R(t) - \langle R(t) \rangle$. The TCF of the birth rate fluctuation vanishes only when the birth process is a Poisson process. As long as the product decay process is a renewal process, Eqs. (1) and (2) hold exactly. For a detailed derivation of Eqs. (1) and (2), see Supplementary Method. Unless there is bursting during the birth process, Eq. (1) reduces to the transient Little's law [18] and Eq. (2) reduces to the CFT [9], owing to the fact that the burst size distribution is a Kronecker delta $\delta_{b1}$. The relation between $\langle \delta R(t_2) \delta R(t_1) \rangle$, the time correlation function (TCF) of the birth rate, and the probability density function of birth time interval is known [19]. Eqs. (1) and (2) reduce to previous results for the $GI^x/G/\infty$ system when both birth and death processes are renewal processes [7].

The contributions to the variance of the product number counting statistics are categorized into three parts: birth rate, product survival probability, and burst size effects. The first term is characterized by the birth rate's mean and TCF. The TCF of the birth rate includes information about the mechanism of the birth process [8,9]. The second term, the product survival probability, can be expressed by a product lifetime distribution provided that the death process is a renewal process. It is necessary to investigate product number counting statistics whose product lifetime distribution is a non-exponential function, because there are cell systems where the product lifetime distribution is a non-exponential function [20-22]. The last term, burst size, is a new factor we introduce in this work to quantify product number fluctuation. The details of burst size effects on product number fluctuation are discussed later in this work.

Our model makes use of two types of vibrant birth processes, a stationary vibrant process and a non-stationary vibrant process. Figure 1(b) provides a visualization of such stationary vibrant-birth processes. The birth rate in Fig. 1(b) fluctuates between on and off states, whose transition rates are respectively given by the constants, $k_{on}$ and $k_{off}$. For the birth model in Fig. 1(b), the mean of the birth rate is constant and its TCF depend only on the difference in time, $\langle \delta R(\tau_2) \delta R(\tau_1) \rangle = f(\tau_2 - \tau_1)$, because the birth process is a stationary process. The mean and TCF of the birth rate for Fig. 1(b) are given as $\langle R \rangle = 5 p_{on}$, $p_{on} \equiv k_{on}/(k_{on} + k_{off})$ and $\langle \delta R(t) \delta R(0) \rangle = e^{-\frac{k_{on}}{p_{on}}t}$.

An example of vibrant non-stationary birth process is given in Fig. 1(c). The birth rate in Fig. 1(c) fluctuates with the stochastic process, $c + w^2$, where $c$ is a positive constant, and $w$ follows the Wiener process $W(t)$, which is a well-known Gaussian, non-stationary stochastic process [23]. Mathematically, $w$ is normally distributed with a mean of zero and variance $at$, $p(w,t) = e^{-\frac{w^2}{2at}} / \sqrt{2\pi at}$. Given that the value of $W(t_0)$ is $w_0$, the transition probability density that the value of $W(t)$ is $w$, is given by $p(w,t|w_0,t_0) = e^{-\frac{(w-w_0)^2}{2a(t-t_0)}} / \sqrt{2\pi a(t-t_0)}$. The mean and the TCF of the birth rate for Fig. 1(c) can then be obtained as $\langle R(t) \rangle = c + at$ and $\langle \delta R(t_2) \delta R(t_1) \rangle = 2a^2 t_1^2$ ($t_2 > t_1$). We model the product lifetime distribution as a gamma distribution in order to calculate product number fluctuation.

Birth rate fluctuation causes product number fluctuation to always appear super-Poisson for the given birth models Figs.1(b)-(c), and the dependence of the product lifetime distribution on product number fluctuation varies with the type of birth process, stationary or non-stationary process. For the two birth rate models considered in the current work, we can see in Figs. 2(a)-(d) the product number counting statistics are always super-Poisson, regardless of the product lifetime fluctuation. We can also see that product number fluctuation is more sensitive to the product lifetime distribution for the stationary vibrant-birth model, Fig. 1(b), than for the non-stationary vibrant-birth model, Fig. 1(c). When the birth rate fluctuates between two states, as shown in Fig. 1(b), the Mandel's Q parameter of the product number always decreases with an increase in the product lifetime fluctuation. This trend is in agreement with the theoretical prediction of reference [9], saturating to a plateau. On the other hand, when the birth rate fluctuation is a nonstationary vibrant process, as shown in Fig. 1(c), product number fluctuation shows a strong time dependence. At short times, product number fluctuation influenced by the product lifetime distribution shows a similar trend to the birth rate model in Fig. 1(b). However, at long times, the Mandel's Q parameter of the product number increases with an increase in the product lifetime fluctuation, which is made clear by the insets in Figs. 2(b) and 2(d). As we can plainly see, the Mandel's Q parameter of the product number diverges as time goes to infinity.

An increase in burst size statistics, mean and Fano factor of burst size $\left(\langle b \rangle, \sigma_b^2/\langle b \rangle \right)$, increases the product number heterogeneity, and a bursting birth process results in a non-zero Mandel's Q parameter at all times. We can clearly see the effect of burst size by comparing Figs. 2(a)-(b) and Figs. 2(c)-(d). In Figs. 2(c) and (d), it is obvious that the product number fluctuation is greater in a bursting birth process than the product number fluctuation in burst-less birth process. In other words, the greater burst size statistics, $\langle b \rangle$ and $\sigma_b^2/\langle b \rangle$, causes the product number heterogeneity to increase; they are positively correlated. We can also make the interesting observation that the Mandel's Q parameter at time zero does not vanish in bursting birth process while it does in a burst-less birth process. For the models considered in this work, the Mandel's Q parameter of the product number at time zero equals $\langle b^2 \rangle / \langle b \rangle - 1$, where $b$ is the burst size. What this means is that the statistics of the burst size, $\langle b^2 \rangle / \langle b \rangle - 1$, can be extracted from the Mandel's Q parameter of the product number at time zero.

In summary, the current work extends the CFT by considering a birth-death model that incorporates bursting birth processes. To this end, we obtain our model's analytic expressions for the product number counting statistics. From our quantitative analysis we prove that the birth rate, product survival probability, and burst size all contribute to the product number fluctuation. Figs. 1 (b)-(c) make it clear that the product number counting statistics appear super-Poisson independent of the product lifetime fluctuation. When the product creation process is a stationary, vibrant process, Fig. 1(b), the product lifetime distribution's influence on product number fluctuation shows a trend in perfect agreement with the theoretical prediction of the CFT [9]; that is, the product number fluctuation decreases as the product lifetime increases. On the other hand, when products are created from the birth rate model in Fig. 1(c), a non-stationary, vibrant process, the product lifetime distribution's influence on the product number fluctuation shows a strong time dependence. At short times, product number fluctuation shows identical trends with the result for the birth model in Fig. 1(b), while we observe the opposite at long times. We also find that burst size is the main contributor to the product number fluctuation. The mean of and the Fano factor of the burst size increases the product number heterogeneity and do not permit the Mandel's Q parameter of the product number to be zero at time zero.

We represent more realistic birth-death models in the current work because we consider both birth rate fluctuation and bursting. Our key findings, namely, Eqs. (1)-(2) and the analytic expressions of product number counting statistics, have broad applications in queueing theory, demography, performance engineering, epidemiology, and biology. For example, Eqs. (1)-(2) can be used to study RNA and protein number, the evolution of bacteria, carriers of a disease in a given population, the number of customers waiting in line at the supermarket, or the number of airplanes waiting the takeoff at the airport. Utilizing Eqs. (1)-(2) as input for the analysis of experimental data probing birth-death processes and product number counting statistics is a topic we leave for future research.

## Acknowledgement

The authors are grateful to Mr. Luke Bates for his careful reading of our manuscript. This work was supported by the Creative Research Initiative Project program (2015R1A3A2066497) funded by the National Research Foundation of the Korean government.

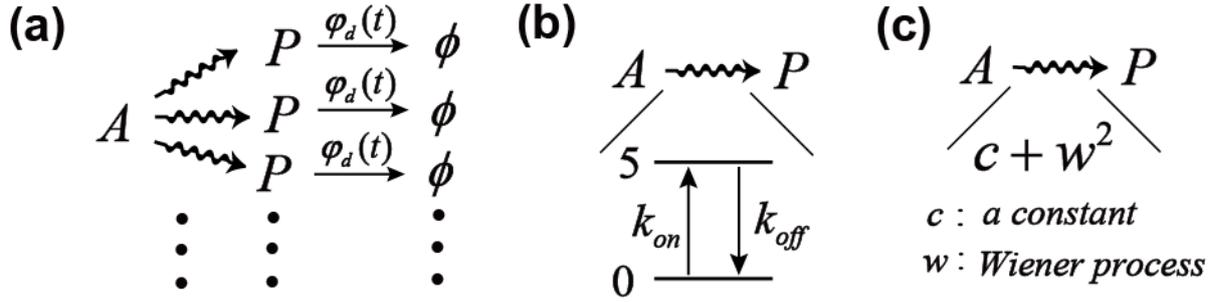

FIG. 1. Birth-death model schematic representation. (a) Wavy arrows represent vibrant birth processes with rate coefficients given by stochastic variables. Vibrant birth processes can be either renewal processes or non-renewal processes. The product, *P*, is created in batches of random size. Every product has an identical lifetime distribution, $\varphi_d(t)$. (b) Vibrant-stationary birth process. The birth rate fluctuates between two states. The value of the birth rate is 5 (0) at the on (off) state. The duration of time the two states is represented by an exponentially distributed random variable; their means are denoted as $k_{on}^{-1}$ and $k_{off}^{-1}$. (c) Vibrant non-stationary birth process. The birth rate is given by the sum of the positive constant, *c*, and the stochastic variable, $w^2$, where *w* fluctuates with the Wiener process.

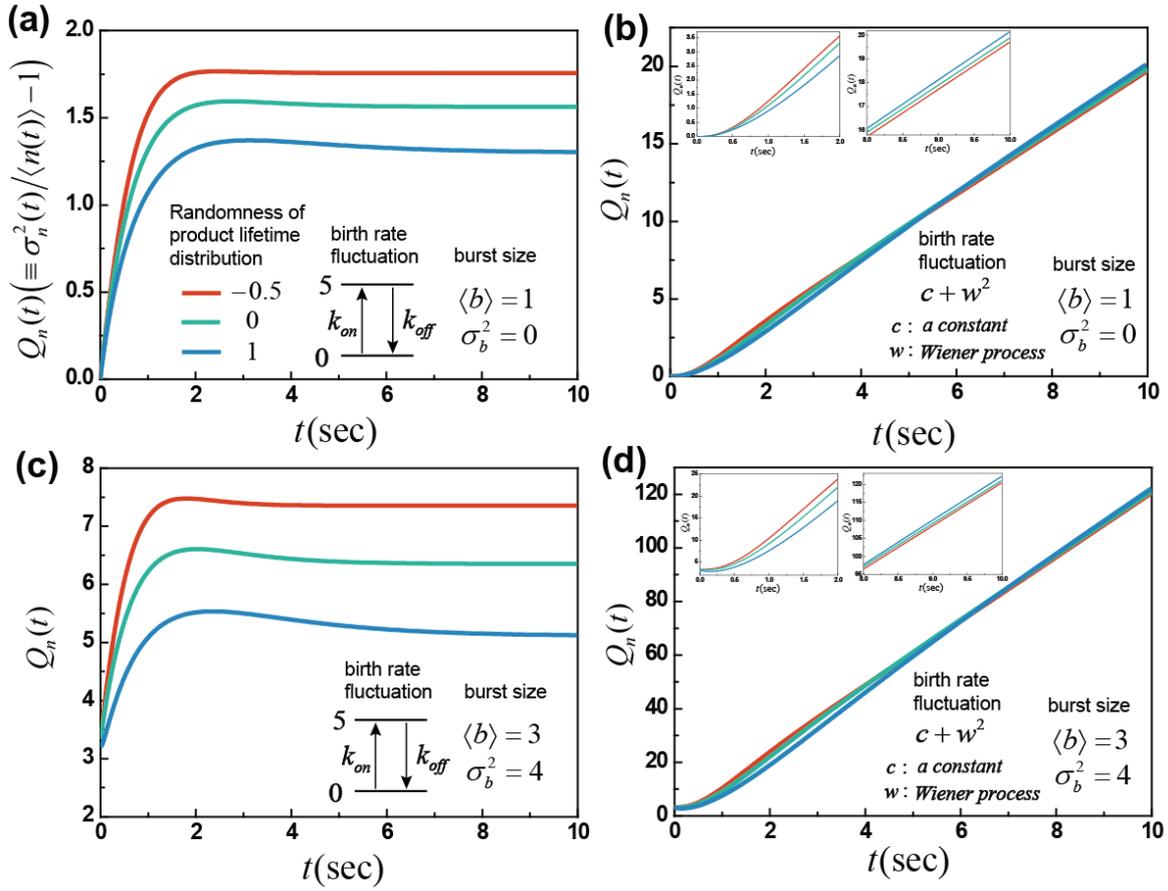

FIG. 2. Mandel's Q parameter of the product number. (a-b) Burst-less birth process. The birth rate fluctuates in accordance with the schemes in Fig 1.(b)-(c). The transition rates $k_{on}$ and $k_{off}$ are set equal to $0.3\,\text{sec}^{-1}$. In case of (b), the constant $c$ is 0.1 and the variable $w$ fluctuates with the Wiener process with a variance of $at$, where $a = 1$. Product lifetime randomness is defined by the relative variance of the product lifetime minus unity, $\left(\langle t_d^2 \rangle - \langle t_d \rangle^2\right)/\langle t_d \rangle^2 - 1$, where $t_d$ is the product lifetime. The mean product lifetime is set equal to 1 sec. (c-d) Bursting birth process, where products are created in random sizes as shown in Fig. 1(a). Birth rate fluctuations of (c)-(d) are respectively identical with (a)-(b). The mean and variance of the burst size are given $\langle b \rangle = 3$ and $\sigma_b^2 = 4$, respectively. Thus, the value of Mandel's Q parameter of product number at time zero in (c)-(d) is $10/3$.

# Supplemental Material

**Product number counting statistics of from stochastic bursting birth-death processes**

Seong Jun Park[1]

[1] National CRI-Center for Chemical Dynamics in Living Cells, Chung-Ang University, Seoul 06974, Korea.

# Table of Contents

**SUPPLEMENTARY METHODS**

**Supplementary Method 1 |** Derivation of Eqs.(1)-(2)

# SUPPLEMENTARY METHODS

**Supplementary Method 1** | Derivation of Eqs.(1)-(2)

In this method, we derive Eqs. (1)-(2) from the main text, which are the mean and variance of a product undergoing a vibrant birth process in batches of a random size and renewal death process. The derivation of Eqs. (1)-(2) is similar to the derivation of the CFT except that we add bursting during the birth process [9].

### 1.1 Mean product number

Let $t_i^c$ denote the time at which the *i*-th group of products is created, $b_i$ its bursting size, and $t_{ij}^d$ the time at which the *j*-th product of the *i*-th group is annihilated. Then, the product number, $n(t)$, at time *t* is given by

$$n(t) = \sum_{i=1}^{\infty} \sum_{j=1}^{b_i} \left( \Theta(t - t_i^c)\left(1 - \Theta(t - t_{ij}^d)\right) \right) \tag{M1-1}$$

where $\Theta(x)$ denotes the Heaviside step function, which assumes 0 for negative *x* but 1 for positive *x*. We additionally assume that $t_i^c$ and $b_i$ are independent of each other. So, we can rewrite the product number as

$$n(t) = \sum_{i=1}^{\infty} \left( \Theta(t - t_i^c) \left( b_i - \sum_{j=1}^{b_i} \Theta(t - t_{ij}^d) \right) \right) \tag{M1-2}$$

In Eq. (M1-1), $\Theta(t - t_{ij}^d)$ can be decomposed into two integrals:

$$\Theta(t - t_{ij}^d) = \int_0^t d\tau \delta(\tau - t_{ij}^d) = \int_0^{t_i^c} d\tau \delta(\tau - t_{ij}^d) + \int_{t_i^c}^t d\tau \delta(\tau - t_{ij}^d) \tag{M1-3}$$

where $\delta(x)$ denotes the Dirac delta function. Because the product decay time, $t_{ij}^d$, is always greater than the *i*-th group of the product creation time, $t_i^c$, the first integral on the right-hand side of Eq. (M1-3) vanishes, i.e., $\int_0^{t_i^c} d\tau \delta(\tau - t_{ij}^d) = 0$. By changing the integration variable from $\tau$ to $\tau' = \tau - t_i^c$ in the remaining integral in Eq. (M1-3), we have

$$\Theta(t - t_{ij}^d) = \int_0^{t-t_i^c} d\tau' \delta\left(\tau' - (t_{ij}^d - t_i^c)\right) = \int_0^{t-t_i^c} d\tau' \delta\left(\tau' - \tau_{ij}\right) \tag{M1-4}$$

where $\tau_{ij} \equiv t_{ij}^d - t_i^c$ denotes the lifetime of the *j*-th product of the *i*-th group. With this identity, $\sum_{i=1}^{\infty} b_i \Theta(t - t_i^c) = \sum_{i=1}^{\infty} b_i \int_0^t d\tau \delta(\tau - t_i^c)$, and Eq. (M1-4) at hand, one can rewrite Eq. (M1-2) as

$$\begin{aligned}n(t) &= \sum_{i=1}^{\infty} \left( b_i \int_0^t d\tau \delta(\tau - t_i^c) - \sum_{j=1}^{b_i} \int_0^t d\tau \delta(\tau - t_i^c) \int_0^{t-t_i^c} d\tau' \delta\left(\tau' - \tau_{ij}\right) \right) \\ &= \sum_{i=1}^{\infty} \left( b_i \int_0^t d\tau \delta(\tau - t_i^c) - \sum_{j=1}^{b_i} \int_0^t d\tau \delta(\tau - t_i^c) \int_0^{t-\tau} d\tau' \delta\left(\tau' - \tau_{ij}\right) \right)\end{aligned} \tag{M1-5}$$

The last equation in Eq. (M1-5) comes from the identity, $\delta(\tau - t_i^c)f(t_i^c) = \delta(\tau - t_i^c)f(\tau)$. By taking the average of Eq. (M1-5) over the distribution of $\{b_i, t_i^c, \tau_{ij}\}$, we obtain

$$\langle n(t) \rangle = \sum_{i=1}^{\infty} \left( \langle b \rangle \int_0^t d\tau \langle \delta(\tau - t_i^c) \rangle - \int_0^t d\tau \langle \delta(\tau - t_i^c) \rangle \int_0^{t-\tau} d\tau' \left\langle \sum_{j=1}^{b_i} \delta(\tau' - \tau_{ij}) \right\rangle \right) \tag{M1-6}$$

Let us now turn our attention to the integrand $\left\langle \sum_{j=1}^{b_i} \delta(\tau' - \tau_{ij}) \right\rangle$ in Eq. (M1-6). Let us regard $\sum_{i=1}^{N} a_i$ as a compound random variable, where $a_i$ and $N$ are random variables. Suppose also that $a_i$ and $N$ are independent of each other. By using the law of total expectation, it is easy to see that $\left\langle \sum_{i=1}^{N} a_i \right\rangle = \langle N \rangle \langle a \rangle$. Likewise, $\left\langle \sum_{j=1}^{b_i} \delta(\tau' - \tau_{ij}) \right\rangle = \langle b \rangle \langle \delta(\tau' - \tau_{ij}) \rangle$. Substituting $\left\langle \sum_{j=1}^{b_i} \delta(\tau' - \tau_{ij}) \right\rangle = \langle b \rangle \langle \delta(\tau' - \tau_{ij}) \rangle$ into Eq. (M1-6), we then obtain

$$\langle n(t) \rangle = \langle b \rangle \sum_{i=1}^{\infty} \int_0^t d\tau \langle \delta(\tau - t_i^c) \rangle \left( 1 - \int_0^{t-\tau} d\tau' \varphi_{ij}(\tau') \right) \qquad \text{(M1-7)}$$

where $\varphi_{ij}(\tau)$ denotes the lifetime distribution, $\langle \delta(\tau - \tau_{ij}) \rangle$, of the *j*-th product of the *i*-th group. Given that every product molecule has the same lifetime distribution, i.e., $\varphi_{ij}(\tau) = \varphi(\tau)$ for any *i* and *j*, we can rewrite Eq.(M1-7) as

$$\langle n(t) \rangle = \langle b \rangle \sum_{i=1}^{\infty} \int_0^t d\tau \langle \delta(\tau - t_i^c) \rangle S(t - \tau) \qquad \text{(M1-8)}$$

where $S(t)$ denotes the survival probability of the product, defined by $S(t) = 1 - \int_0^t d\tau \varphi(\tau)$.

The term $\sum_{i=1}^{\infty} \delta(t - t_i^c)$ in Eq.(M1-8) is the production creation rate when products do not decay and are not created in a batch of a random size, and we denote this production creation rate by $R(t) = \sum_{i=1}^{\infty} \delta(t - t_i^c)$. By substituting $\langle R(t) \rangle = \sum_{i=1}^{\infty} \langle \delta(t - t_i^c) \rangle$ into Eq. (M1-8), one obtains Eq. (1) from the main text.

### 1.2 Product number variance

Deriving Eq. (2) is similar Eq. (1)'s derivation. From Eq. (M1-2), we obtain the following equation for $n^2(t)$:

$$\begin{aligned} n^2(t) &= \sum_{i=1}^{\infty} \sum_{k=1}^{\infty} \Theta(t - t_i^c) \Theta(t - t_k^c) \left( b_i - \sum_{j=1}^{b_i} \Theta(t - t_{ij}^d) \right) \left( b_k - \sum_{l=1}^{b_k} \Theta(t - t_{kl}^d) \right) \\ &= \sum_{i=1}^{\infty} \sum_{k=1}^{\infty} b_i b_k \Theta(t - t_i^c) \Theta(t - t_k^c) - 2 \sum_{i=1}^{\infty} \sum_{k=1}^{\infty} b_i \Theta(t - t_i^c) \Theta(t - t_k^c) \sum_{l=1}^{b_k} \Theta(t - t_{kl}^d) \\ &+ \sum_{i=1}^{\infty} \sum_{k=1}^{\infty} \Theta(t - t_i^c) \Theta(t - t_k^c) \sum_{j=1}^{b_i} \Theta(t - t_{ij}^d) \sum_{l=1}^{b_k} \Theta(t - t_{kl}^d) \end{aligned} \qquad \text{(M1-9)}$$

The first summation on the right-hand side of Eq. (M1-9) can be decomposed into the following

two summations:

$$\sum_{i=1}^{\infty}\sum_{k=1}^{\infty}b_ib_k\Theta(t-t_i^c)\Theta(t-t_k^c) = \sum_{i=1}^{\infty}b_i^2\Theta(t-t_i^c) + \sum_{i=1}^{\infty}\sum_{\substack{k=1\\k\neq i}}^{\infty}b_ib_k\Theta(t-t_i^c)\Theta(t-t_k^c) \quad \text{(M1-10)}$$

where the first term on the right-hand side results from summing over the terms with $i=k$. Similarly, one can further decompose the remaining two summations of Eq. (M1-9). We can then further decompose the second term Eq. (M1-9)'s right-hand side:

$$\begin{aligned}&\sum_{i=1}^{\infty}\sum_{k=1}^{\infty}b_i\Theta(t-t_i^c)\Theta(t-t_k^c)\sum_{l=1}^{b_k}\Theta(t-t_{kl}^d)\\ &= \sum_{i=1}^{\infty}b_i\Theta(t-t_i^c)\sum_{l=1}^{b_i}\Theta(t-t_{il}^d) + \sum_{i=1}^{\infty}\sum_{\substack{k=1\\k\neq i}}^{\infty}b_i\Theta(t-t_i^c)\Theta(t-t_k^c)\sum_{l=1}^{b_k}\Theta(t-t_{kl}^d)\end{aligned} \quad \text{(M1-11)}$$

For the last term of Eq. (M1-9) we obtain

$$\begin{aligned}&\sum_{i=1}^{\infty}\sum_{k=1}^{\infty}\Theta(t-t_i^c)\Theta(t-t_k^c)\sum_{j=1}^{b_i}\Theta(t-t_{ij}^d)\sum_{l=1}^{b_k}\Theta(t-t_{kl}^d)\\ &= \sum_{i=1}^{\infty}\Theta(t-t_i^c)\left(\sum_{j=1}^{b_i}\Theta(t-t_{ij}^d)\right)^2 + \sum_{i=1}^{\infty}\sum_{\substack{k=1\\k\neq i}}^{\infty}\Theta(t-t_i^c)\Theta(t-t_k^c)\sum_{j=1}^{b_i}\Theta(t-t_{ij}^d)\sum_{l=1}^{b_k}\Theta(t-t_{kl}^d)\end{aligned} \quad \text{(M1-12)}$$

By substituting Eqs. (M1-10)-(M1-12) into the right-hand side of Eq. (M1-9) and taking the average of $n^2(t)$ over $\{t_i^c, t_k^c\}$, $\{t_{ij}^d, t_{kl}^d\}$, and $\{b_i, b_k\}$, we find

$$\begin{aligned}\langle n^2(t)\rangle &= \sum_{i=1}^{\infty}\langle b^2\rangle\langle\Theta(t-t_i^c)\rangle - 2\sum_{i=1}^{\infty}\langle\Theta(t-t_i^c)\rangle\left\langle b_i\sum_{l=1}^{b_i}\Theta(t-t_{il}^d)\right\rangle\\ &+ \sum_{i=1}^{\infty}\langle\Theta(t-t_i^c)\rangle\left\langle\left(\sum_{j=1}^{b_i}\Theta(t-t_{ij}^d)\right)^2\right\rangle\\ &+ \sum_{i=1}^{\infty}\sum_{\substack{k=1\\k\neq i}}^{\infty}\left(\begin{aligned}&\langle b\rangle^2\langle\Theta(t-t_i^c)\Theta(t-t_k^c)\rangle - 2\langle b\rangle\langle\Theta(t-t_i^c)\Theta(t-t_k^c)\rangle\left\langle\sum_{l=1}^{b_k}\Theta(t-t_{kl}^d)\right\rangle\\ &+\langle\Theta(t-t_i^c)\Theta(t-t_k^c)\rangle\left\langle\sum_{j=1}^{b_i}\Theta(t-t_{ij}^d)\right\rangle\left\langle\sum_{l=1}^{b_k}\Theta(t-t_{kl}^d)\right\rangle\end{aligned}\right)\end{aligned} \quad \text{(M1-13)}$$

We need to identify $\left\langle N\sum_{i=1}^{N} a_i \right\rangle$ and $\left\langle \left(\sum_{i=1}^{N} a_i\right)^2 \right\rangle$, the statistics of a compound random variable, where $a_i$ and $N$ are random variables, independent of each other. By the law of total expectation, we obtain

$$\left\langle N\sum_{i=1}^{N} a_i \right\rangle = \left\langle N^2 \right\rangle \left\langle a \right\rangle \tag{M1-14}$$

we can rewrite $\left\langle \left(\sum_{i=1}^{N} a_i\right)^2 \right\rangle$ as

$$\left\langle \left(\sum_{i=1}^{N} a_i\right)^2 \right\rangle = Var\left(\sum_{i=1}^{N} a_i\right) + E(N)^2 E(a)^2 \tag{M1-15}$$

where each $Var(z)$ and $E(z)$ respectively denote the variance and the mean of the random variable z and we used $E\left(\sum_{i=1}^{N} a_i\right) = E(N)E(a)$. Before proceeding, let us briefly discuss the law of total variance, which is relevant to the rest of the derivation. The law of total variance states that if $x$ and $y$ are random variables on the same probability space, and the variance of $Y$ is finite, then

$$Var(y) = E\big(Var(y|x)\big) + Var\big(E(y|x)\big) \tag{M1-16}$$

where $Var(y|x)$ and $E(y|x)$ respectively denote the conditional variance and mean of $y$, given $x$. Using the law of total variance on Eq. (M1-15), the first term of the right-hand side becomes

$$Var\left(\sum_{i=1}^{N} a_i\right) = E\left[Var\left(\sum_{i=1}^{N} a_i \middle| N\right)\right] + Var\left(E\left[\sum_{i=1}^{N} a_i \middle| N\right]\right)$$

$$= E\left[Var\left(\sum_{i=1}^{N} a_i\right)\right] + Var(NE(a)) \quad \text{(M1-17)}$$

$$= E(NVar(a)) + E(a)^2 Var(N)$$

$$= Var(a)E(N) + E(a)^2 Var(N)$$

The independence of $a_i$ and $N$ confirms the second line in Eq. (M1-17). Substituting Eq. (M1-17) into Eq. (M1-15), we obtain

$$\left\langle \left(\sum_{i=1}^{N} a_i\right)^2 \right\rangle = \langle \delta a^2 \rangle \langle N \rangle + \langle a \rangle^2 \langle N^2 \rangle \quad \text{(M1-18)}$$

Equations (M1-14) and (M1-18) can serve as proof of Eq. 2 from the main text.

We are able to rewrite Eq. (M1-13) by applying Eqs. (M1-14) and (M1-18) to $\left\langle b_i \sum_{l=1}^{b_i} \Theta(t-t_{il}^d) \right\rangle$ and $\left\langle \left(\sum_{j=1}^{b_i} \Theta(t-t_{ij}^d)\right)^2 \right\rangle$. We then find

$$\langle n^2(t) \rangle = \langle b^2 \rangle \sum_{i=1}^{\infty} \langle \Theta(t-t_i^c) \rangle - 2\langle b^2 \rangle \sum_{i=1}^{\infty} \langle \Theta(t-t_i^c) \rangle \langle \Theta(t-t_{ij}^d) \rangle$$

$$+ \sum_{i=1}^{\infty} \langle \Theta(t-t_i^c) \rangle \left( \langle (\delta\Theta(t-t_{ij}^d))^2 \rangle \langle b \rangle + \langle \Theta(t-t_{ij}^d)^2 \rangle \langle b^2 \rangle \right) \quad \text{(M1-19)}$$

$$+ \sum_{\substack{i=1 \\ k \neq i}}^{\infty} \sum_{k=1}^{\infty} \left( \begin{array}{l} \langle b \rangle^2 \langle \Theta(t-t_i^c)\Theta(t-t_k^c) \rangle - 2\langle b \rangle \langle \Theta(t-t_i^c)\Theta(t-t_k^c) \rangle \left\langle \sum_{l=1}^{b_k} \Theta(t-t_{kl}^d) \right\rangle \\ + \langle \Theta(t-t_i^c)\Theta(t-t_k^c) \rangle \left\langle \sum_{j=1}^{b_i} \Theta(t-t_{ij}^d) \right\rangle \left\langle \sum_{l=1}^{b_k} \Theta(t-t_{kl}^d) \right\rangle \end{array} \right)$$

where $\langle b^m \rangle$ is the m-th moment of the burst size $b$. The only remaining obstacle is $\langle \delta\Theta(t-t_{ij}^d)^2 \rangle$ in Eq. (M1-19). This problem is easily overcome by using $\langle \Theta(t-t_{ij}^d)^2 \rangle = \langle \Theta(t-t_{ij}^d) \rangle$. We immediately find

$$\langle \delta\Theta(t-t_{ij}^d)^2 \rangle = \langle \Theta(t-t_{ij}^d) \rangle \left(1 - \langle \Theta(t-t_{ij}^d) \rangle\right) \tag{M1-20}$$

Given that every product molecule has the same lifetime distribution, $\varphi(\tau)$, we can represent $\langle \Theta(t-t_{ij}^d) \rangle$ in terms of $\varphi(\tau)$ by using similar argument used to obtain Eq. (M1-7) from (M1-6),

$$\langle \Theta(t-t_{ij}^d) \rangle = \int_0^t d\tau \varphi(\tau) \equiv r_d(t) \tag{M1-21}$$

where $r_d(t)$ is the probability that the annihilation reaction occurs within time $t$. It is simple to then find the relationship between the reaction probability, $r_d(t)$, and the survival probability of the product, $S(t)$, that is, $S(t) = 1 - r_d(t)$. We now use the equation,

$\left\langle \sum_{l=1}^{b_k} \Theta(t-t_{kl}^d) \right\rangle = \langle b \rangle r_d(t)$, and Eqs. (M1-19)-(M1-21), obtaining

$$\langle n^2(t) \rangle = \langle b^2 \rangle \int_0^t d\tau \langle R(\tau) \rangle \left(1 - 2r_d(t-\tau)\right) + \langle b \rangle \int_0^t d\tau \langle R(\tau) \rangle r_d(t-\tau) S(t-\tau)$$
$$+ \langle b^2 \rangle \int_0^t d\tau \langle R(\tau) \rangle r_d(t-\tau)^2 \tag{M1-22}$$
$$+ 2\langle b \rangle^2 \sum_{i=1}^{\infty} \sum_{\substack{k=1 \\ k \neq i}}^{\infty} \int_0^t d\tau_2 \int_0^{\tau_2} d\tau_1 \langle \delta(\tau_2 - t_i^c)\delta(\tau_1 - t_k^c) \rangle S(t-\tau_1) S(t-\tau_2)$$

where $R(t) = \sum_{i=1}^{\infty} \delta(t-t_i^c)$ is the product creation rate without bursting. Noting the definition of the product creation rate without bursting, $R(t)$, given in $R(t) = \sum_{i=1}^{\infty} \delta(t-t_i^c)$, we identify $\sum_{i=1}^{\infty} \sum_{\substack{k=1 \\ k \neq i}}^{\infty} \langle \delta(\tau_1 - t_i^c)\delta(\tau_2 - t_k^c) \rangle$ as the time correlation function of the product creation rate, i.e.

$$\langle R(\tau_2) R(\tau_1) \rangle = \sum_{i=1}^{\infty} \sum_{\substack{k=1 \\ k \neq i}}^{\infty} \langle \delta(\tau_2 - t_i^c) \rangle \langle \delta(\tau_1 - t_k^c) \rangle \tag{M1-23}$$

Substituting $r_d(t) = 1 - S(t)$ and Eq. (M1-23) into Eq. (M1-22), we obtain

$$\langle n^2(t) \rangle = \langle n(t) \rangle + \left( \langle b^2 \rangle - \langle b \rangle \right) \int_0^t d\tau \langle R(\tau) \rangle S(t-\tau)^2 \\ + 2\langle b \rangle^2 \int_0^t d\tau_2 \int_0^t d\tau_1 \langle R(\tau_2) R(\tau_1) \rangle S(t-\tau_1) S(t-\tau_2) \quad \text{(M1-24)}$$

Subtracting $\langle n(t) \rangle^2$, with $\langle n(t) \rangle$ given in Eq. (1) from the main text, from Eq. (M1-24), we obtain Eq. (2) from the main text. Equations (1) and (2) hold even when the birth process is a non-stationary process.

When the birth process is a stationary process, we have $\langle R(\tau_2) \rangle = \langle R(\tau_1) \rangle = \langle R \rangle$ and $\langle R(\tau_2) R(\tau_1) \rangle = \langle R(\tau_2 - \tau_1) R(0) \rangle$. Substituting these equations into Eqs. (1)-(2), we obtain

$$\langle n(t) \rangle = \langle b \rangle \langle R \rangle \int_0^t d\tau S(\tau) \quad \text{(M1-25)}$$

$$\sigma_n^2(t) = \langle n(t) \rangle + \left( \langle b^2 \rangle - \langle b \rangle \right) \langle R \rangle \int_0^t d\tau S(\tau)^2 \\ + 2\langle b \rangle^2 \int_0^t d\tau_2 \int_0^{\tau_2} d\tau_1 \langle \delta R(\tau_2 - \tau_1) \delta R(0) \rangle S(t-\tau_1) S(t-\tau_2) \quad \text{(M1-26)}$$

where $\langle \delta R(t) \delta R(0) \rangle$ designates $\langle R(t) R(0) \rangle - \langle R \rangle^2$. As stated previously, $S(t)$ denotes the survival probability of a product molecule, and $\langle b^m \rangle$ is the m-th moment of burst size $b$. Equations (M1-25)-(M1-26) are applicable to any stationary-birth process.